\documentclass[aps,pre,amsmath,amsfonts,floatfix,superscriptaddress]{revtex4}

\usepackage{graphicx}
\usepackage{feynmf}
\font\msytw=msbm9 scaled\magstep1

\let\a=\alpha \let\b=\beta    \let\d=\delta 
     
\let\m=\mu             \let\p=\pi   
    \let\f=\varphi 
   
\let\G=\Gamma \let\D=\Delta  \let\Th=\Theta 
    \let\Si=\Sigma     
 
\let\io=\infty
\def\ie{{i.e. }}\def\eg{{e.g. }}

\def\FF{{\cal F}}

 \def\xx{{\bf x}} \def\yy{{\bf y}} 

\def\uu{{\bf u}}

\def\erf{\text{erf}}

\def\to{\rightarrow}
\def\la{\left\langle}
\def\ra{\right\rangle}

\def\RRR{\hbox{\msytw R}}

\newcommand{\beq}{\begin{equation}}
\newcommand{\eeq}{\end{equation}}
\newcommand{\wh}{\widehat}

\begin{document}

\title{Amorphous packings of hard spheres in large space dimension}

\author{Giorgio Parisi\footnote{giorgio.parisi@roma1.infn.it}}

\affiliation{Dipartimento di Fisica, Universit\`a di Roma ``La Sapienza'', 
P.le A. Moro 2, 00185 Roma, Italy}

\affiliation{INFM -- CRS SMC, INFN, Universit\`a di Roma ``La Sapienza'', 
P.le A. Moro 2, 00185 Roma, Italy}

\author{Francesco Zamponi\footnote{francesco.zamponi@phys.uniroma1.it}}

\affiliation{Laboratoire de Physique Th\'eorique de l'\'Ecole Normale Sup\'erieure,
24 Rue Lhomond, 75231 Paris Cedex 05, France}

\date{\today}

\begin{abstract}
In a recent paper we derived an expression for the replicated free energy
of a liquid of hard spheres based on the HNC free energy functional.
An approximate equation of state for the glass and an estimate
of the random close packing density were obtained in $d=3$.
Here we show that the HNC approximation is not needed: the same expression
can be obtained from the full diagrammatic expansion of the replicated
free energy. Then, we
consider the asymptotics of this expression when the space dimension
$d$ is very large. 
In this limit, the entropy of the hard sphere liquid has been computed exactly.
Using this solution, we derive asymptotic
expressions for the glass transition density and for the random close packing density
for hard spheres in large space dimension.
\end{abstract}


\maketitle

\begin{fmffile}{diagrammi1}

\newcommand{\diauno}{
\fmfframe(2,0)(0,0){
\begin{fmfchar*}(8,5)
  \fmfleft{a}
  \fmfright{d}
  \fmf{plain}{a,d}
  \fmfdot{a,d}
\end{fmfchar*}}
}

\newcommand{\diadue}{
\begin{fmfchar*}(8,5)
  \fmfleft{a,q}
  \fmfright{b,q}
  \fmftop{c}
  \fmf{plain}{a,b,c,a}
  \fmfdot{a,b,c}
\end{fmfchar*}}

\newcommand{\diatre}{
\begin{fmfchar*}(8,5)
  \fmfleft{a,b}
  \fmfright{d,c}
  \fmf{plain}{a,b,c,d,a}
  \fmfdot{a,b,c,d}
\end{fmfchar*}}

\newcommand{\diaquattro}{
\begin{fmfchar*}(8,5)
  \fmfleft{a,b}
  \fmfright{d,c}
  \fmf{plain}{a,b,c,d,a}
  \fmf{plain}{a,c}
  \fmfdot{a,b,c,d}
\end{fmfchar*}
}

\newcommand{\diacinque}{
\fmfframe(0,0)(2,0){
\begin{fmfchar*}(8,5)
  \fmfleft{a,b}
  \fmfright{d,c}
  \fmf{plain}{a,b,c,d,a}
  \fmf{plain}{a,c}
  \fmf{plain}{b,d}
  \fmfdot{a,b,c,d}
\end{fmfchar*}
}}

\newcommand{\diasei}{
\begin{fmfchar*}(8,5)
  \fmfleft{a,q}
  \fmfright{b,q}
  \fmftop{c}
  \fmf{dashes}{a,b,c,a}
\end{fmfchar*}}

\newcommand{\diasette}{
\begin{fmfchar*}(8,5)
  \fmfleft{a,q}
  \fmfright{b,q}
  \fmftop{c}
  \fmf{dashes}{a,c,b}
  \fmf{photon}{a,b}
\end{fmfchar*}}

\section{Introduction}

The study of amorphous packings of hard spheres is relevant for a large class
of problems, including liquids, glasses, colloidal dispersions, granular
matter, powders, porous media, etc. 
\cite{Be83,SK69,Fi70,Be72,Ma74,Po79,Al98,SEGHL02,To95,RT96,Sp98,Torquato,Luca05}.
Nevertheless, the question whether a glass transition exists for a liquid of
identical hard spheres in finite dimension is still open.

Recently, a quantitative description of the glass transition in structural glasses has been obtained
by means of the replica trick \cite{MP99,MP99b,MP00,CMPV99,CFP98,PZ05}.  The latter is applied in
the context of a variational approximation - the hypernetted chain (HNC) approximation - that leads
to a suitable free energy functional for simple liquids \cite{Hansen,MH61}.  This method was
successfully applied to Lennard-Jones systems \cite{MP99,MP99b,MP00,CMPV99} and, more recently, to
hard spheres in space dimension $d=3$ \cite{CFP98,PZ05}.  Quantitative estimates of the glass
transition temperature (or density) and of the equation of state of the glass have been obtained in
this way.  Moreover, for hard spheres an estimate of the random close packing density, \ie the
maximum density of the amorphous configurations of the system, has been obtained as the value of the
density where the pressure of the glass diverges \cite{PZ05}.

In this approximation the glass transition turns out to be similar to the 1-step replica symmetry
breaking (1RSB) transition that happens in a class of mean-field spin glass models and indeed the
replica strategy described above was inspired by the exact solution of these models
\cite{MPV87,KTW87b,GM84,Mo95}.  However, for finite dimensional models with short range interactions
the picture emerging from the replica-HNC approach should be modified by non-perturbative activated
processes (for a detailed discussion see \eg \cite{XW01,BB04,Fr05}).  The results obtained in $d=3$
have often been considered as a kind of ``mean-field'' approximation.

Activated processes should become less relevant on increasing the space dimension.  It is then
natural to ask whether this approximation describes better and better the true properties of the
system (eventually becoming exact) when the space dimension is large as it is well known that the
mean field approximation becomes exact in the $d\to\io$ limit.  Moreover, the study of sphere
packings in large space dimension is relevant for information theory and lot of effort has been
devoted to finding the densest packing for $d\to \io$ \cite{ConwaySloane,Rogers}.  Despite this
effort, only some not very restrictive bounds have been obtained, and it is still unclear whether
the densest packings for $d\to \io$ are amorphous or crystalline.

In this paper we will: {\it i)} show that in $d\to\io$ the replica approach predicts
the existence of a glass transition density $\rho_K$ and compute the value of this density;
{\it ii)} compute the maximum density $\rho_c$ of the amorphous packings (or glass states) in
$d \to \io$. Unfortunately the value of $\rho_c$ we find is within the current bounds for crystalline
packings so we cannot address the problem whether the densest packings are amorphous or not.

The paper is organized as follows: first we briefly outline the method, without entering in
details as they are discussed in \cite{PZ05}; then we present an improvement of the theory
of \cite{PZ05}, \ie we show how to construct the small cage expansion without using 
the HNC approximation; finally we discuss the limit $d\to\io$ of the theory.

\section{The method}

In the replica method one considers a liquid made of $m$ copies of the original system, with the
constraint that each atom of a given replica must be close to an atom of the other $m-1$ replicas,
\ie that the replicated liquid must be made of {\it molecules} of $m$ atoms, each one belonging to a
different replica.  It was shown that this trick allows to compute all the properties of the glass,
including the size of the cages, the vibrational entropy, etc., provided one is able to perform the
analytical continuation to real $m \leq 1$, see \cite{MP99,MP99b,MP00,Mo95} for a detailed
discussion.  Thus the problem is to compute the free energy of the replicated system for integer $m$
and to continue the expression for real $m$.

This strategy has been first tested on mean field spin glass models \cite{Mo95,MP99b}, and then
applied to systems of particles interacting through a Lennard-Jones like potential
\cite{MP99,MP00,CMPV99}.  To compute the replicated free energy for a system of particles, the idea
was to start from the standard HNC free energy for a molecular liquid~\cite{MH61} and expand it in a
power series of the ``cage radius'', that represents the amplitude of the vibrations of the
particles in the glass state~\cite{MP99b}.  The method was successful but could not be extended
straightforwardly to hard spheres, because at some stage it was assumed that vibrations were
harmonic, an approximation that clearly breaks down for hard core potentials.

The small cage expansion of the replicated HNC free energy was worked out in \cite{PZ05}.
It was shown that the theory is in reasonable quantitative agreement with numerical data
even if the HNC approximation is a very poor description of the hard spheres liquid.
In next section we will show that the result of \cite{PZ05} for the replicated
free energy, obtained starting from the HNC free energy functional, can be derived 
from the full diagrammatic expansion, without the need of neglecting any class of diagrams.

\section{Small cage expansion}

Our approach will be based on the standard diagrammatic approach (virial expansion). It seems to us
that similar  results 
could also be obtained by a direct computation, however we think that it is more instructive to use 
the more familiar diagrammatic approach. 

We start from the diagrammatic expansion of the canonical free energy
as a function of the single-molecule density $\rho(\xx)$ and of the interaction
potential between molecules. It can be obtained from the
grancanonical partition function of the system via a Legendre 
transform~\cite{MH61}.
Calling $\xx \equiv (x_1,\ldots,x_m)$ the coordinate of a molecule,
the expression for the replicated free energy functional is \cite{Hansen,MH61}
(setting from now on $\b = 1$):
\beq
\Psi[\rho(\xx),f(\xx,\yy)] = \int d\xx \rho(\xx) [\log \rho(\xx) -1 ] - \left[
\diauno + \diadue + \diatre + \diaquattro + \diacinque
+ \cdots
\right]
\eeq
where the black circles represent a function $\rho(\xx)$ and the lines represent
a function $f(\xx,\yy) = e^{-\f(\xx,\yy)} - 1 $, and $\f(\xx,\yy) = \sum_a \f(|x_a-y_a|)$ is 
the interaction potential between two molecules, with $\f(r)=\io$ for $r < D$ and $0$ otherwise,
the usual hard sphere interaction. 
The function
$f(\xx,\yy)$ is equal to $-1$ if $|x_a - y_a| < D$ for at least one pair $a$ of
spheres, and vanishes otherwise.
The information on the inter-replica interaction that is used to build
molecules is encoded in the function $\rho(\xx)$.

The generic diagram of the series expansion of $\Psi$ represents an integral of the form
\beq\label{dia3}
\diadue
=\frac{1}{S} \int d\xx_1 d\xx_2 d\xx_3 \rho(\xx_1)\rho(\xx_2)\rho(\xx_3) 
f(\xx_1,\xx_2) f(\xx_2,\xx_3) f(\xx_3,\xx_1) \ .
\eeq
where $S$ is the symmetry number of the diagram \cite{Hansen}.

For the one-molecule density we make the {\it ansatz}
\beq\label{rhoGauss}
\rho(\xx) = \rho \int dX \prod_a \frac{e^{-\frac{(x_a-X)^2}{2A}}}{(\sqrt{2\p A})^d}
= \rho \int dX \, \wh \rho(\uu)
\ , \hskip20pt \wh \rho(\uu) = \prod_a \frac{e^{-\frac{(u_a)^2}{2A}}}{(\sqrt{2\p A})^d}
\eeq
where $\rho=N/V$ is the number density of molecules,
$X$ is the center of mass of the molecule and $u_a \equiv x_a-X$.
This allows to
compute exactly the ideal gas term of the free energy:
\beq\label{gasperfetto}
\frac{1}{N} \int d\xx \rho(\xx) [\log \rho(\xx) -1 ] =
\log \rho - 1 + \frac{d}{2}(1-m)\log(2\pi A) + \frac{d}{2} (1-m-\log m) \ .
\eeq
We want to perform a power series expansion in $\sqrt{A}$ of the interaction term.

At the zeroth order,
the functions $\wh\rho(\uu)$ are products of delta functions, $\wh\rho(\uu) = \prod_a \d (u_a)$, 
so that all
the coordinates $x_a$ are equal to $X$. This gives a contribution
\beq
\diasei
=\frac{\rho^3}{S} \int dX_1 dX_2 dX_3 F(X_1,X_2) F(X_2,X_3) F(X_3,X_1) \ .
\eeq
where $F(X,Y) = -\theta(D-|X-Y|)$ is represented by a dashed line 
and a vertex $i$ without a black dot represents
the constant $\rho$ and an integration over the corresponding $X_i$.
This is exactly the contribution of the
diagram to the usual non-replicated free energy.
Thus, at the zeroth order the molecular free energy simply reduces to the
free energy $\FF[\rho,F(r)]$ of a liquid made by the center of masses of the 
molecules, with density $\rho$ and an hard core interaction.

To discuss the first order correction in $\sqrt{A}$ let us first insert
Eq.~(\ref{rhoGauss}) into (\ref{dia3}):
\beq
\diadue
= \frac{\rho^3}{S} \int dX_1 dX_2 dX_3 d\uu_1 d\uu_2 d\uu_3 
\wh\rho(\uu_1)\wh\rho(\uu_2)\wh\rho(\uu_3) 
f(X_1+\uu_1,X_2+\uu_2) f(X_2+\uu_2,X_3+\uu_3) f(X_3+\uu_3,X_1+\uu_1) \ ,
\eeq
using the notation $X+\uu = (X+u_1, \ldots, X+u_m)$ and
\beq
f(X_1+\uu_1,X_2+\uu_2) = e^{-\sum_a \f(X_1-X_2+u_{1a}-u_{2a})} -1 \ .
\eeq
For small $A$ the $u$ are small too, $u \sim \sqrt{A}$.
The function $f$ above has the property
that, if $|X_1-X_2|$ differs from $D$ by a quantity $\gg \sqrt{A}$, it is independent of $\uu_1, \uu_2$: 
in fact it is a constant equal to $-1$ if $|X_1-X_2|<D$ and $0$ otherwise.
That is, if $|X_1-X_2|$ is not close to $D$, one has 
$f(X_1+\uu_1,X_2+\uu_2) = F(X_1,X_2)$ and the diagram does not give any contribution
apart from the zeroth order discussed above (recall that $\int d\uu \wh\rho(\uu) = 1$).

Thus, the corrections in $\sqrt{A}$ due to a given diagram 
come from the region of the
integration space where at least two of the coordinates $X_i$, $X_j$ connected by a
link have distance $D$, $|X_i - X_j | \sim D + O(\sqrt{A})$. Let us call such a
pair a ``singular pair'', and the link connecting them as a ``singular link''.
The regions of the integration space of the $X_i$ where $n$ pairs 
$X_i$, $X_j$ are singular have a 
total volume $\sim (\sqrt{A})^n$.
Thus the only contribution at order $\sqrt{A}$ comes from
$n=1$, \ie only one singular pair.

In all the non-singular links we can replace $f(X_i+\uu_i,X_j+\uu_j) 
\rightarrow F(X_i,X_j)$ and we get
\beq\begin{split}
\diadue
&=\frac{1}{S} \int d\xx_1 d\xx_2 d\xx_3 \rho(\xx_1)\rho(\xx_2)\rho(\xx_3) f(\xx_1,\xx_2) f(\xx_2,\xx_3)
f(\xx_3,\xx_1) \\ &=
\diasei
+ \frac{3\rho^3}{S} \int dX_1 dX_2 dX_3 
\left[\int d\uu_1 d\uu_2 \wh \rho(\uu_1)\wh \rho(\uu_2) f(X_1+\uu_1,X_2+\uu_2) - F(X_1,X_2) \right] 
F(X_2,X_3)
F(X_1,X_3)  \\ &=
\diasei
+ \diasette
\ ,
\end{split}
\eeq
where the wiggly line represents the function
\beq\label{Qdef}
Q(X_1,X_2) = \int d\uu_1 d\uu_2 \wh\rho(\uu_1)\wh\rho(\uu_2) f(X_1+\uu_1,X_2+\uu_2) - F(X_1,X_2) \ ,
\eeq
which is different from $0$ only if $|X_1-X_2| \sim D + O(\sqrt{A})$.
Thus for each diagram the correction is obtained by replacing one $F$-link
with one $Q$-link in all the possible non-equivalent ways. This is equivalent
to taking the derivative of the zeroth order diagram with respect to $F(X_1,X_2)$,
multiplying by $Q(X_1,X_2)$ and integrating over $X_1$, $X_2$.
Summing the contribution of all the diagrams we then obtain the first order correction
to the replicated free energy:
\beq\label{correction1}
\D \Psi[\rho(\xx),f(\xx,\yy)] =  \int dX_1 dX_2 \frac{\d \FF}{\d
  F(X_1,X_2)} Q(X_1,X_2) \ .
\eeq
where the derivative with respect to $F(X_1,X_2)$ of the free energy is taken
at constant $\rho$. Recalling that the canonical free energy is the Legendre
transform of the grancanonical free energy, it easy to prove that \cite{Hansen,MH61}
\beq\label{11}
 \frac{\d \FF}{\d F(X_1,X_2)} = - \frac{\d \log Z_{GC} }{\d F(X_1,X_2)} =
- \frac{\rho^2}{2} Y(X_1,X_2) \ ,
\eeq
where the function $Y(r) \equiv e^{\f(r)} G(r)$ is continuous for hard spheres.

The calculation of the function $Q(X_1,X_2) = Q(|X_1-X_2|)$ was already done
in \cite{PZ05}
and is reported in Appendix~\ref{app:A}:
\beq\label{12}
Q(r) = 2 \sqrt{A} Q_m \Si_d(D) \d(r-D) \ ,
\eeq
where $Q_m$ is an analytic function of $m$, 
with $Q_m \sim 0.638 (1-m) + o((1-m)^2)$, see Appendix~\ref{app:A},
and $\Si_d(D)=\frac{ 2 \pi^{d/2} D^{d-1}}{ \G(d/2)}$ is the surface
of a $d$-dimensional hypersphere of radius $D$.

Substituting Eq.s~(\ref{11}), (\ref{12}) in Eq.~(\ref{correction1}), and adding the correction
coming from the ideal gas term, Eq.~(\ref{gasperfetto}), we obtain
\beq\label{Fcorrect}
\begin{split}
\Psi[\rho(\xx),f(\xx,\yy)] &=\FF(\rho) 
 + N \left[ \frac{d}{2}(1-m)\log(2\pi A) + \frac{d}{2} (1-m-\log m)\right]
- \rho^2 \sqrt{A} Q_m \Si_d(D) V \int_0^\io dr \d(r-D)
Y(r) \\ & =  \FF(\rho) 
 + N \left[ \frac{d}{2}(1-m)\log(2\pi A) + \frac{d}{2} (1-m-\log m)\right]
- N \rho \sqrt{A} Q_m \Si_d(D) Y(D) \ ,
\end{split}
\eeq
where $\FF(\rho)$ is the free energy of the non-replicated liquid and
$Y(D)=G(D)$ is the value of the pair correlation function at contact.
This is the same result obtained in \cite{PZ05} from the expansion of the HNC
free energy, but here it was obtained
{\it without need of the HNC approximation}. This is important because the HNC
approximation gives very poor results for hard spheres systems.

Recalling that $\FF(\rho) = - N S(\rho)$,
where $S(\rho)$ is the equilibrium entropy per particle of the liquid,
we obtain the replicated free energy by optimizing with respect to $A$ in
Eq.~(\ref{Fcorrect}):
\beq\label{repF}
\begin{split}
&\Phi ( m , \rho)= \frac1N \min_A \Psi[\rho(\xx),f(\xx,\yy)] = 
-S(\rho) + \frac{d}2(1-m) \log[2\p A^*] + \frac{d}2(m-1-\log m) \ , \\
&\sqrt{A^*(m)} = \frac{1-m}{Q_m} \frac{D}{\rho V_d(D) Y(D)}
\end{split}
\eeq
where
$V_d(D) =\frac{ 2 \pi^{d/2} D^d}{ d \G(d/2)}$ 
is the volume of a sphere of radius $D$.
The function $A^*(m)$ has the meaning of a ``cage radius'' as discussed above.
This result holds in any space dimension $d$. 

In $d=3$ we used the Carnahan-Starling expression \cite{Hansen} for the entropy $S(\rho)$, 
which reproduces very well the numerical data for the equation of state of the
hard sphere liquid.
This was done in \cite{PZ05} on a phenomenological ground, but is fully justified by the present
derivation in which no reference to the HNC approximation is done.
It was shown that Eq.~(\ref{repF}) predicts a glass transition density $\rho_K$ which is
in good agreement with numerical results. Moreover, in the glass phase the
cage radius decreases with the density and reaches $0$ at a value $\rho_c$ which
is then the maximum allowed density for an amorphous state, \ie the random
close packing density. For $\rho \to \rho_c$ the pressure of the
glass diverges. 
It was also shown that the average number of neighbors of a given sphere at
$\rho_c$ is equal to $z=2d$.
A detailed discussion and a comparison with numerical
data can be found in \cite{PZ05}. Here we will discuss the predictions of Eq.~(\ref{repF})
in the limit $d\to\io$.

\section{Entropy of the liquid in the limit of large dimension}

The problem of computing the entropy of the hard sphere liquid for
$d\to \io$ was addressed in \cite{FP99,PS00}, where the same result was obtained
in two independent ways.
In \cite{FP99} it was shown that the ring diagrams dominate the virial series order
by order in $\rho$ for large $d$, and the entropy was computed by a resummation of these
diagrams;
in \cite{PS00} simple equations for the pair correlation function $g(r)$ were introduced,
and solved in the limit $d\to\io$.
In both cases it was found that $S(\rho)$ is given by the ideal gas term plus
the first virial correction (\ie by the Van der Waals equation).

The equations introduced in \cite{PS00} are indeed the minimal requirements for a
pair distribution function $g(r)$. If $D$ is 
the sphere diameter, $\rho=N/V$ is the density, $h(r) = g(r) - 1$ and $h(q)$ is
its Fourier transform, one has:
\beq\label{simple}
\begin{cases}
g(r) \geq 0  \ , \\
g(r) = 0 \text{ for } r \geq D \ , \\ 
S(q)=1+\rho h(q) \geq 0 \ ,
\end{cases}
\eeq
The first condition comes from the fact that $g(r)$ is a probability (the
probability of finding a particle at distance $r$ given that there is a
particle in the origin), the second from the fact that two particles cannot
be at distance smaller than $D$ (due to the hard core repulsion). The third
condition is easily obtained by proving that $S(q) = 1 + \rho h(q) = \la | \rho_q
|^2 \ra$, where $\rho_q$ is a Fourier component of the density fluctuations \cite{Hansen}.

In \cite{PS00} a solution of Eq.s~(\ref{simple}) for $d \to \io$ was found. 
We set the sphere diameter $D=1$ (defining $V_d\equiv V_d(1)$ and $Y\equiv
Y(1)$) and following \cite{PS00}
we define $d=2N+3$, the reduced density $\overline \rho \equiv \rho V_d$ and 
\begin{equation}
\rho_1 \equiv
\frac{\log \overline \rho}{N} \equiv \frac{2 \log \overline \rho}{d-3} \, . 
\end{equation}
Note that $\overline\rho$ is related to the packing fraction $\phi\equiv \rho V_d(1/2)$, \ie to the
fraction of volume covered by the spheres, by $\phi=2^{-d} \overline\rho$.
It is found that for $\overline \rho \leq 1$ the solution of Eq.s~(\ref{simple}) is
simply $g(r) = \theta(r-1)$, while for $\overline \rho > 1$ it has the form
$g(r) = \theta(r-1) \{1 + \exp[- N h_1(r) ] \}$ with $h_1(r) > 0$, 
\ie it is given by the step function
plus an exponentially small correction which can be explicitly computed, see
\cite{PS00}.
The pressure is then found to be:
\beq \label{PPS}
\frac{P}{\rho} = 1+ \begin{cases}
\frac12 \overline \rho \hskip80pt \text{for } \overline \rho \leq 1 \ , \\
\frac12 \overline \rho (1 +e^{-2 N K(p_c)}) \hskip10pt \text{for } \overline \rho > 1 \ ,
\end{cases}
\eeq
where 
\beq
K(p) = \log (1+\sqrt{1-p^2}) - \sqrt{1-p^2} - \log p \ ,
\eeq
and $p_c$ the solution of
\beq
\phi_0(p) = \log (1+\sqrt{1-p^2}) - \sqrt{1-p^2} + 1 -\log 2 = \rho_1 \ .
\eeq
It turns out that $K(p_c)>0$ for $0\leq \rho_1 < \phi_0(1) = 1-\log 2 =
0.3068..$. Essentially the pressure is given by the ideal gas contribution
plus the first virial correction, with another correction which is
exponentially small up to $\rho_1 = 1-\log 2$. Above the latter value a solution
of Eq.s~(\ref{simple}) could not be found.

These results are strongly consistent with the results of \cite{FP99} where it was
shown that the resummation of the ring diagrams gives exactly Eq.~(\ref{PPS}) up to
$\rho_1 = 1 -\log 2$. At this value of the density a pole develops that seems to correspond
to a liquid instability (the {\it Kirkwood instability} \cite{FP99}). 

It not clear if this instability implies that there are no solutions of
Eq.s~(\ref{simple}) (or of the HNC equations) for high
densities, or it simply implies that one has to look for more complex
solutions.
Although the question is very interesting from the mathematical point of view (maybe also in
relation to the problem of finding the most dense lattices \cite{PP}), this
question is not 
physically relevant 
in this contest. We will show later that
the glass transition indeed preempts this instability that is therefore in a non-physical region of
the density: this system becomes unstable toward replica symmetry breaking before reaching the
Kirkwood instability. 

Using the exact relation
\beq\label{exact}
\frac{P}{\rho} = 1 +\frac12 \overline \rho Y = - \rho \frac{d S}{d\rho} \ ,
\eeq
it turns out that
\beq\label{Fdgrande}
S(\rho) = 1-\log \rho - 
\begin{cases}
\frac12 \overline \rho \hskip100pt \text{for } \overline \rho \leq 1 \ , \\
\frac12 \overline \rho (1 + L(p_c) e^{-2 N K(p_c)}) \hskip10pt \text{for } \overline \rho > 1 \ ,
\end{cases}
\eeq
where, recalling that $\overline\rho \frac{dp_c}{d\overline \rho} = \frac1N\frac{dp_c}{d\rho_1} = \frac{1}{N\phi_0'(p_c)}$,
$L(p)$ is such that
\beq
L(p) + \frac{L'(p) - 2NK'(p) L(p)}{N \phi_0'(p)} = 1 \hskip10pt 
\Rightarrow \hskip10pt L(p) \sim \frac{1}{1 - 2 K'(p)/\phi_0'(p)} 
= \frac{1}{2 \frac{1+\sqrt{1-p^2}}{p^2} - 1 } \ .
\eeq
Again, up to exponentially small corrections, the entropy of the liquid is
given by the ideal gas term plus the first virial correction.
Similarly, by comparing Eq.s~(\ref{exact}) and (\ref{PPS}) we find that $Y =
1$ up to exponentially small corrections.

\section{Glass transition in large space dimension}

To locate the glass transition we substitute Eq.~(\ref{Fdgrande})
in Eq.~(\ref{repF}) and
compute the equilibrium complexity \cite{MP99,MP99b,MP00,Mo95,PZ05}:
\beq
\Si(\rho) = \left. m^2 \frac{\partial (\Phi/m)}{\partial m} \right|_{m=1} =
S(\rho) - \frac{d}2 \log[2\pi A^*(1)] = S(\rho) - S_{vib}(\rho) \ ,
\eeq
where $S_{vib}(\rho) \equiv \frac{d}2 \log[2\pi A^*(1)]$ is
the vibrational contribution to the entropy of the liquid;
the glass transition density (or Kauzmann density) $\rho_K$ is defined by $\Si(\rho)=0$. 
Note that the cage radius is, from Eq.~(\ref{repF}), $\sqrt{A} \propto
\overline\rho^{-1}$; thus for $\rho_1 \leq 0$ it is exponentially large in $N$ and the small
cage expansion does not make sense.
\begin{figure}[t]
\includegraphics[width=8cm]{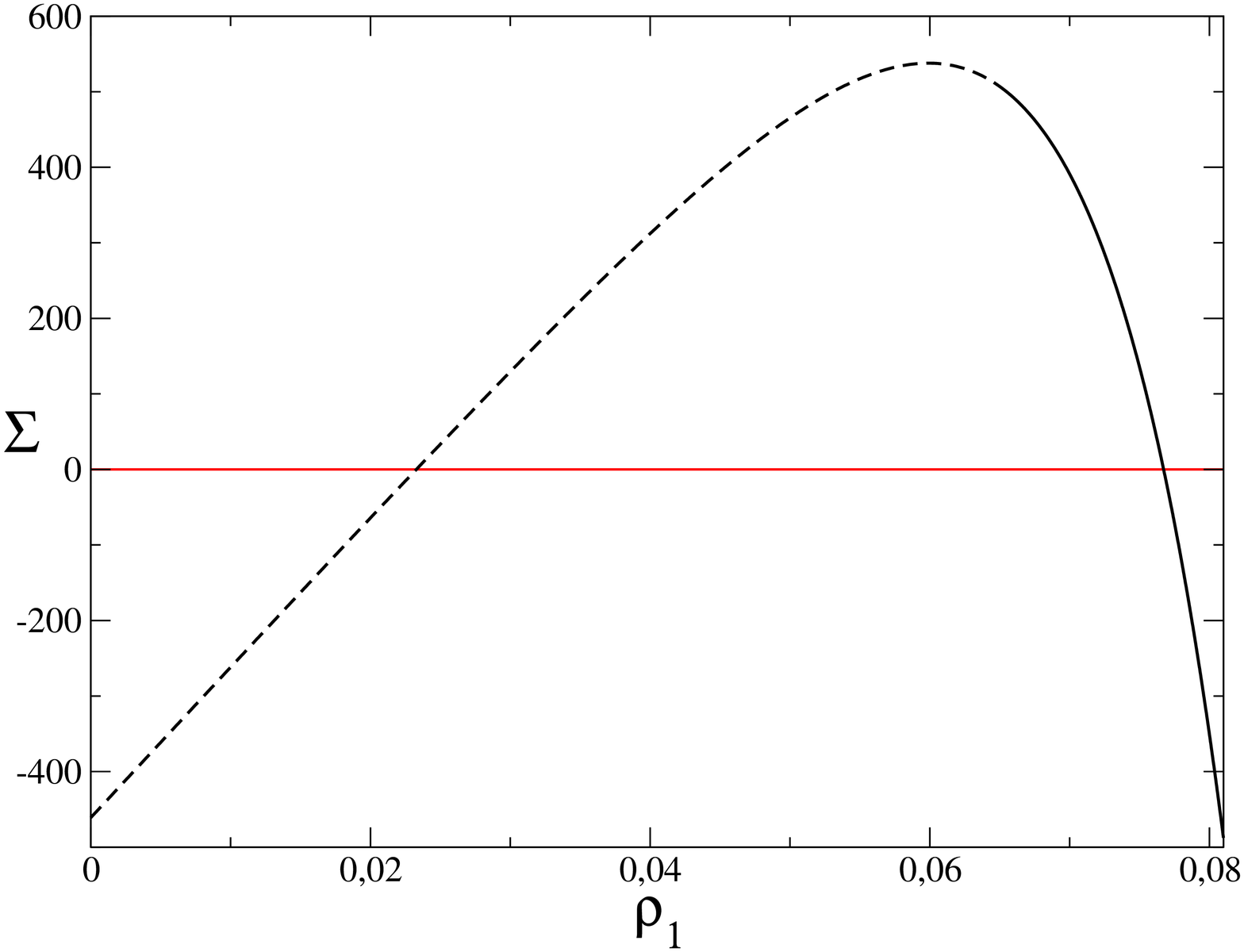}
\caption{The function $\Si(\rho_1)$, Eq.~(\ref{sir}), for $N=100$.
Our computation is based on Eq.~(\ref{repF}) which comes from an expansion in
powers of the cage radius and 
is valid only close to the Kauzmann density where $\Si = 0$.
The dashed part of the curve is then unphysical. Unfortunately, the exact point where
Eq.~(\ref{repF}) breaks down could not be estimated. Thus in the figure the
point where the curve becomes dashed has been chosen arbitrarily.
}
\label{fig1}
\end{figure}
For $\rho_1 > 0$ we get, neglecting the terms related to $K(p_c)$ which are exponentially
small in $N$, and using $V_d \sim \left(\frac{e \pi}{N}\right)^N \frac{1}{N
  \sqrt{2\pi N}}$,
\beq\label{Sr}
\begin{split}
&S(\rho) =  1 -\log\rho -\frac12 \overline \rho \sim
1 - N\rho_1 + N \log (e\p) - N\log N - \log (\sqrt{2\p} N^{3/2}) - \frac12 e^{N\rho_1} \ , \\ 
&S_{vib}(\rho) = d \log \frac{\sqrt{2 \p}}{0.638} 
+d \log [\overline \rho (1 + e^{-2 N K(p_c)}) ] \sim
2 N \log \frac{\sqrt{2 \p}}{0.638} - (2 N +3) N \rho_1 \ , 
\end{split}
\eeq
and the equation for $\rho_K$ is, neglecting for simplicity the terms growing
slower than $N$, and defining $\a = 2 \log  \frac{\sqrt{2 \p}}{0.638} -
\log(e\p) \sim 0.59$,
\beq\label{sir}
0=\Si(\rho_1)  \sim 
 - N \a - N \log N + 2 N \rho_1 (N + 1) 
-\frac12 e^{N\rho_1} 
\ .
\eeq
A plot of $\Si(\rho_1)$ is reported in Fig.~\ref{fig1}. 
Let us first neglect the terms proportional to $N$: then
the equation $\Si=0$ gives
\beq\label{Si0}
N \log N = 2 N^2 \rho_1 - \frac12 e^{N \rho_1} \ ,
\eeq
and a solution is simply $N\rho_1 = \frac12 \log N$. However the derivative of
$\Si$ evaluated in this solution is
\beq
\Si'(\rho_1) = 2 N^2 - \frac12 N e^{N\rho_1} 
\eeq 
and is positive, thus this value corresponds to the unphysical zero of $\Si$
(see Fig.~\ref{fig1}).
The maximum of $\Si$ is given by $\Si'=0$, \ie $N\rho_1 = \log 4N$.
Thus we look for a solution of Eq.~(\ref{Si0}) of the form
\beq
e^{N\rho_1}= q N \ ,
\eeq
and we expect that $q$ diverges with $N$.
Substituting in Eq.~(\ref{Si0}) we get
\beq
q = 4 \log ( q \sqrt{N} ) = 2 \log N + 4 \log q \ ,
\eeq
which is solved iteratively by
\beq\label{q0}
q = 2 \log N + 4 \log ( 2 \log N + 4 \log q ) =
2 \log N + 4 \log [2 \log N + 4 \log ( 2 \log N + 4 \log q ) ] = \ldots  \ \ .
\eeq
Thus the value of the Kauzmann density is
\beq\begin{split}
&\overline \rho_K = 2 N \log N + O(N\log \log N) \ , \\
&\rho_{1K} = \frac{1}{N} \log(2N \log N + O(N\log \log N)) \ .
\end{split}
\eeq

\subsection{Correction to the Kauzmann density}

For future convenience it is useful to compute the correction
to $\overline \rho_K$ due to the terms $O(N)$ we discarded in Eq.~(\ref{sir}).
The full Eq.~(\ref{sir}) is 
\beq\label{fullSi0}
- N \log N + 2 N^2 \rho_1 - \frac12 e^{N\rho_1} = N \a -2 N \rho_1 \ .
\eeq
We look again for a solution $e^{N \rho_1} = N q$ with
$q$ large. Then we obtain
\beq
4 \log q +2 \log N - q = 2 \a - \frac{4}N \log (q N) \ .
\eeq
The right hand side is $o(1)$ while the left hand side diverges so the leading
solution is $q = q_0$ with $q_0$ given by Eq.~(\ref{q0}). We look for a solution
$q = q_0 + q_1$ with $q_1\ll q_0$: then the left hand side gives
\beq
4 \log ( q \sqrt{N}) - q = 4 \log \left(1 + \frac{q_1}{q_0}\right)
 -q_1 \sim q_1 \left( \frac{4}{q_0} -1\right) \ ,
\eeq
then
\beq
 q_1 \left( \frac{4}{q_0} -1\right) = 2 \a + O(N^{-1} \log N) \ ,
\eeq
and finally as $q_0 \sim \log N$
\beq
q_1 = -2 \a + O((\log N)^{-1}) \ .
\eeq
The result for $\overline \rho_K$ is then
\beq\label{rK}
\overline \rho_K = N q_0 - 2 N \a + O\left(\frac{N}{\log N}\right) = N q_0 - 1.18 N \ ,
\eeq
with $q_0$ given by Eq.~(\ref{q0}).

\subsection{Random close packing density}

For $\overline \rho > \overline \rho_K$ the system is in the glass phase and the value $m^*$
that optimizes the free energy (per replica) is $m^* < 1$ \cite{MP99,MP99b,MP00,Mo95,PZ05}.
$m^*$ is the solution of $\Si(\rho,m) = 0$, where
\beq\begin{split}
\Si(\rho,m) = m^2 \frac{\partial (\Phi/m)}{\partial m} &=
S(\rho) - N \log( 2 \pi) + (2 N+3)N \rho_1 - N m -Nm(1-m) \frac{Q_m'}{Q_m} \\ &+ N \log m - 2 N\log (1-m)
+2 N \log Q_m \ .
\end{split}
\eeq
As $Q_m  \sim \sqrt{\pi/4m}$ for $m\to 0$ (see Appendix~\ref{app:A}), 
one can see from Eq.~(\ref{repF})
that the cage radius $A^* \to 0$ for $m \to 0$. The random
close packing density $\overline\rho_c$ is then the value of $\overline \rho$ at which $m^* = 0$ \cite{PZ05}.
Then to compute $\overline \rho_c$ we have to solve $\Si(\overline \rho,0) = 0$. Using
$Q_m \sim \sqrt{\pi/4m}$,
\beq
\Si(\overline\rho,0) = S(\rho) + (2 N+3) N \rho_1 + N \left( \frac12 - 3 \log 2  \right) \ .
\eeq 
The condition $\Si(\overline\rho,0) = 0$, using Eq.~(\ref{Sr}), is similar to Eq.~(\ref{fullSi0}):
\beq
- N \log N + 2 N^2 \rho_1 - \frac12 e^{N\rho_1} = N \a' -2 N \rho_1 \ ,
\eeq
with $\a' = 3 \log 2 -\frac12 -\log (e \p) \sim -0.56$. The solution is given
by Eq.~(\ref{rK}) with $\a \to \a'$:
\beq\begin{split}
&\overline \rho_c =  N q_0 - 2 N \a' + O\left(\frac{N}{\log N}\right) = N q_0 + 1.12 N 
= \overline \rho_K + 2.3 N \ , \\
&\rho_{1c} = \frac{1}{N} \log (  N q_0 + 1.12 N ) \ .
\end{split}
\eeq

\subsection{Lindemann ratio}

To check the consistency of the small cage expansion, it is interesting to estimate the
Lindemann ratio in the glass phase, when $\overline \rho \sim 2N \log N \sim d \log d$.
The Lindemann ratio $L$ for a given solid phase is the ratio between the typical amplitude 
of vibrations around the equilibrium positions and the mean interparticle distance. In our
framework it can be defined as
\beq
L \equiv \rho^{1/d} \sqrt{A}
\eeq
so that, using $\sqrt{A} \sim 1 / \overline\rho$ from Eq.~(\ref{repF}),
and $\overline \rho = \rho V_d \sim d \log d$, one has
\beq
L \sim \frac{1}{\sqrt{d} \log d} \ll 1 \ ,
\eeq
which is consistent with the assumption that vibrations are very small.

\subsection{Is the densest packing amorphous in large $d$?}

We can compare our prediction for the maximum density of amorphous packings
$\overline\rho_c \sim d \log d$ with the best available bounds on the density of crystalline
packings. The corresponding packing fraction scales as $\phi_c \sim 2^{-d} d \log d$.
Unfortunately, the best lower bound for periodic packings is the Minkowski bound $\overline\rho \sim 1$,
while the best upper bound is the Blichfeldt's one, $\overline\rho \sim 2^{d/2}$
\cite{ConwaySloane,Rogers}. Our result for $\overline\rho_c$ lies between these bounds so
we cannot give an answer to the question whether the densest packings of hard spheres
in large $d$ are amorphous or crystalline. Hopefully better bounds on the density
of crystalline packings will address this question in the future. However it is difficult to escape 
to the impression that the values of the densities of crystalline laminated lattices \cite{ConwaySloane}
up to $d$=50 suggest that there are lattices where $\rho_1$ goes to a non-zero limit for infinite $d$. 
It is however quite possible that this is a preasymptotic effect. It would be very interesting to
find the density of laminated lattices in larger dimensions. 

It would be also interesting to find out
the maximum value of the density for which the inequalities (\ref{simple}) have a solution, allowing
also delta functions in $h(p)$: those equations are valid also for regular lattices; however the
problem is hard and it is not clear how to attack the problem mathematically (it is a linear
programming problem in an infinite dimensional space also for finite dimensions).

\section{Conclusions}

Making use of the results of \cite{FP99,PS00,PZ05} we showed that 
the small cage expansion of the replicated free energy
in large space dimension 
predicts a 1-step replica symmetry breaking transition. We obtained  
the asymptotic behavior of the glass transition density and of the random close packing density
as $d \to \io$. 
It is worth to note that all the phenomena related to the glass
transition happen is a region of densities which is very close to $\overline\rho=1$, in the sense
that $\rho_1 = (d/2)^{-1} \log \overline\rho$ is bounded by $o(\log d / d)$, while
the Kirkwood instability \cite{FP99,PS00}
happens for $\rho_1 = 1-\log 2 = 0.30\ldots$, \ie 
well beyond the interesting range of values of $\rho_1$. In other words, the region of
densities where the Kirkwood instability happens is never reached due to the glass transition
happening at lower density.

It would be interesting to estimate the corrections at finite $d$ to the
asymptotic expressions. For the simplified model~(\ref{simple}) the corrections
are exponentially small in $d$ as found in \cite{PS00}. 
It is then reasonable that the exact expression for the liquid entropy as
well differs from the asymptotic solution of~(\ref{simple}) by exponentially
small terms, and the same might happen for the replicated free energy.
If this is the case, the results we presented in this paper should be exact
for $d\to\io$. We hope that future work will address this point.

\acknowledgments

We are pleased to thank G.~Biroli, J.-P.~Bouchaud, and S.~Franz for useful discussions.
One of us (F.Z.) benefited a lot from a discussion at the end of his seminar in LPTHE,
Jussieu and wishes to thank all the participants for their comments and suggestions.

This work has been supported by the Research Training Network STIPCO (HPRN-CT-2002-00319).

\appendix
\section{The function $Q(r)$}
\label{app:A}

The function $Q(r)$ has been defined in Eq.~(\ref{Qdef}) as:
\beq
Q(r) = \int d\uu_1 d\uu_2 \wh\rho(\uu_1)\wh\rho(\uu_2) f(r+\uu_1,\uu_2) - F(r) \ ,
\eeq
where $F(r) = -\theta(D-|r|)$ and $f(r+\uu_1,\uu_2) = -1 + \prod_a e^{-\f(r+u_{1a}-u_{2a})}$.
Recalling that $\int d\uu \wh\rho(\uu) = 1$ we can rewrite
\beq
Q(r) = \int d\uu_1 d\uu_2 \wh\rho(\uu_1)\wh\rho(\uu_2) \prod_a e^{-\f(r+u_{1a}-u_{2a})}
- \theta(|r|-D) = F_0(r)^m - \theta(|r|-D) \ ,
\eeq
where
\beq
F_0(r) = \int du_1 du_2 \frac{e^{-\frac{(u_1)^2+(u_2)^2}{2A}}}{(\sqrt{2\p A})^d} \theta(|r+u_1-u_2|-D) =
\int du \frac{e^{-\frac{u^2}{4A}}}{(\sqrt{4\p A})^d} \theta(|r+u|-D) \ .
\eeq
From the expressions above we expect that $Q(r)$ is non zero only if $|r| \sim D$.

Let us first discuss the expansion of $Q(r)$ in $d=1$. Defining
\beq
\erf(t) \equiv \frac{2}{\sqrt{\p}} \int_0^t dx \, e^{-x^2} \ , \hskip20pt
\Th(t) = \frac{1}{2} [ 1 + \erf(t) ] = \frac{1}{\sqrt{\pi}} \int_{-\io}^t  dx \, e^{-x^2} \ ,
\eeq
we have
\beq
\int_{-\io}^\io du \, \frac{e^{-\frac{u^2}{4A}}}{\sqrt{4\p A}} \,
\theta(r+u-D) = \Th\left(\frac{r-D}{\sqrt{4A}}\right) \ .
\eeq
Note that for small $A$, $u$ is small too, and as $|r| \sim D$, the sign of $r+u$ is the same as
the sign of $r$. Thus we can write $\theta(|r+u|-D) \sim \theta(r+u-D) + \theta(-r-u-D)$
and
\beq
\label{F0}
F_0(r) =\Th\left(\frac{r-D}{\sqrt{4A}}\right)
+ \Th\left(-\frac{r+D}{\sqrt{4A}}\right) \sim  \Th\left(\frac{|r|-D}{\sqrt{4A}}\right)  \ .
\eeq
where in the last step we  neglected contributions of order $\exp(-D^2/A)$ for $A \rightarrow 0$,
and 
\beq
Q(r) = \left[ \Th\left(\frac{|r|-D}{\sqrt{4A}}\right) \right]^m - \theta(|r|-D) \ .
\eeq

From the expression above it is easy to see that $Q(r)$ is non zero only if $|r|-D\sim \sqrt{A}$, as
expected. This is because the function $\Th(t)$ is exponentially close to $\theta(t)$ if $t$ is large.
Given a smooth function $f(|r|)$, one has $\int dr Q(r) f(r) \propto \sqrt{A} f(D)$ for small $\sqrt{A}$,
\ie $Q(r) \propto \sqrt{A} \, \d(|r|-D)$. We have now to compute the proportionality factor.
Defining the reduced variable $t= (r-D)/\sqrt{4A}$, we have:
\beq
\int_0^\io dr Q(r) =
2\sqrt{A} \int_{-\frac{D}{\sqrt{4A}}}^\io dt [ \Th(t)^m - \theta(t) ] \equiv 2 \sqrt{A} Q(A) \sim
2 \sqrt{A} Q_m + o(\sqrt{A}) \ ,
\eeq
where the function $Q_m$ is the limit for $A \rightarrow 0$ of $Q(A)$ and is given by
\beq
Q_m = \int_{-\io}^\io dt \, [ \Th(t)^m - \theta(t) ] \ .
\eeq
It is easy to show that $Q_m$ is a finite and smooth function of $m$ for
$m \neq 0$, that
\beq
\begin{split}
&Q_m = (1-m) Q_0 + O[(m-1)^2] \ , \\
&Q_0 = -\int_{-\io}^\io dt \, \Th(t) \log \Th(t) \sim 0.638 \ ,
\end{split}
\eeq
and that $Q_m$ diverges as $Q_m \sim \sqrt{\p/4m}$ for $m \rightarrow 0$.
Thus we get the result $Q(r) = 2 \sqrt{A} Q_m \d(|r|-D)$ in $d=1$.

In dimension $d>1$ we have, recalling that $F_0(r)$ is rotationally invariant,
for $R=|r|$:
\beq
\label{DDDD}
\int dr Q(r) = \int d r \, [ F_0(r)^m - \theta(|r|-D) ] = 
\Omega_d \int_0^\io dR \, R^{d-1} \, [ F_0(R)^m - \theta(R-D) ] \ ,
\eeq
where $\Omega_d$ is the solid angle in $d$ dimension, $\Omega_d=2\pi^{d/2}/\G(d/2)$.
The function $F_0(R)$ can be written as
\beq
F_0(R) = \int d u \, \frac{e^{-\frac{u^2}{4A}}}{(\sqrt{4\p A})^d} \theta(|R \widehat i + u|-D) \ ,
\eeq 
where $\widehat i$ is the unit vector e.g. of the first direction in $\RRR^d$. For small $\sqrt{A}$, 
the $u$ are small too, and the function $\theta(|r \widehat i + u|-D)$ is constant along the directions
orthogonal to $\widehat i$. 
Thus we can show that the integral over the variables $u_\m, \m\neq 1$ is equal to $1$ up to corrections
of the order of $\exp(-D/\sqrt{A})$.
We finally get:
\beq
\label{5D}
F_0(R) = \int_{-\io}^\io du_1 \, \frac{e^{-\frac{u_1^2}{4A}}}{\sqrt{4\p A}} \theta(|R + u_1|-D)
=  \Th\left(\frac{R-D}{\sqrt{4A}}\right)   \ ,
\eeq 
as in the one dimensional case. Again, the function $F_0(R)^m - \theta(R-D)$ is large only for
$R - D \sim \sqrt{A}$ so at the lowest order we can replace $R^{d-1}$ with $D^{d-1}$
in Eq.~(\ref{DDDD}). Finally, using that $F_0(R)$ is given by Eq.~(\ref{5D}) as in $d=1$, we get
\beq
\int dr Q(r) = \Omega_d D^{d-1} \int_0^\io dR \, [ F_0(R)^m - \theta(R-D) ] = \Si_d(D) 2 \sqrt{A} Q_m \ ,
\eeq
where $\Si_d(D)$ is the surface of a $d$-dimensional sphere of radius $D$,
$\Si_d(D) = \Omega_d D^{d-1}$, \ie in any dimension $d$ we have
\beq
Q(r) = 2 \sqrt{A} Q_m \d(|r|-D) \ .
\eeq

\end{fmffile}

\end{document}